# Multi-objective optimization of the coiled carbon nanotubes regarding their mechanical performance: A reactive molecular dynamics simulation study

Ehsan Shahini [a,*], Fazel Rangriz [b], Ali Karimi Taheri [a]

[a] Department of Material Science and Engineering, Sharif University of Technology, Azadi Ave. Tehran, Iran, PO Box: 11155-9466

[b] Department of Electronic Systems, Norwegian University of Science and Technology, NTNU, NO-7491 Trondheim, Norway

## Abstract

Coiled Carbon Nanotubes (CCNTs) are increasingly set to become a vital factor in the new generation of nanodevices and energy-absorbing materials due to their outstanding properties. In the following work, the multi-objective optimization of CCNTs is applied regarding their mechanical performances. Apart from finding the best trade-off between conflicting mechanical properties (e.g. yield stress and yield strain), the optimization enables us to find the astonishing CCNTs concerning their stretchability. To the best of our knowledge, these structures have not been recognized before, both experimentally and computationally. Several highly accurate analytical equations are derived by insights from the findings of multi-objective optimization and fitting a theoretical model to the results of Molecular Dynamics (MD) simulations. The structures resulted from optimizations are highly resilient because of two distinct deformation mechanisms depending on the dimensions of CCNTs. For small CCNTs, extraordinary extensibility is mainly contributed by buckling and nanohinge-like deformation with maintaining the inner coil diameter, whereas for large CCNTs this is accomplished by the creation of a straight CNT-like structure in the inner-edge of the CCNT with a helical graphene ribbon twisted around it. These findings would shed light on the design of CCNT based mechanical nanodevices.

[*] Corresponding author. Tel: +98362527627. E-mail: Ehsanshahiny@gmail.com (Ehsan Shahini)

# 1. Introduction

The helical shape is a prevalent configuration in the universe from spiraling galaxies to protein α-helix and DNA double helix. Therefore, it is not surprising that this should also be a common motif observed in carbon nanostructures [1]. Because of their unique 3D helical morphology, relatively high electrical conductivity [2,3], large surface area [4], high-performance electromagnetic wave absorption [5,6], and superelasticity [7–13], CCNTs are applicable in a variety of fields such as electrocatalyst for fuel cells [14–18], supercapacitor electrodes [19,20], reinforcement [21,22], biological sensors [23], hydrogen storage materials [24,25], and chiral catalysts [4]. In mechanics, the ability of CCNTs to elastically sustain loads at large deflections allows them to store or absorb significant amounts of strain energy. This should render composites enhanced by helical CNTs relevant where energy-absorbing characteristics are desired [26]. Thus, to better understand their applications, it is essential to study the CCNT's mechanical behavior. To discover the mechanical properties of CCNTs, a large amount of pioneering experimental and theoretical research was performed [8,16,35–38,27–34].

Experimentally, Volodin et al. [30] evaluated a Young's module of about 0.7 TPa for helical CNT with a coil diameter of 170 nm using atomic force microscopy (AFM). The spring constant and maximum strain of a double wall CCNT with 126 nm tubular diameter was determined by Chen et al. [39]. They clamped the CCNT between the two cantilevers of AFM and stretched up to 42% strain. Their results showed a nonlinear spring-like stretching response with a spring constant of 0.12 N/m. Hayashida et al. [27] by using a manipulator-equipped scan electron microscopy (SEM), reported that elastic modulus of CCNTs varies from 0.04 to 0.13 TPa for coil radius ranging from 72 to 415 nm. Poggi et al. [29] evaluated the compressive strength of CCNTs with different length, coil diameter, and the number of walls and identified a buckling behavior of multi-walled CCNTs using in situ AFM.

Theoretically, the tensile response of CCNTs of various diameters was investigated at different temperatures [8]. The results of this research have verified that the tension force was reduced by raising the temperature and reducing the diameter of CCNTs. Ghaderi and Haji Esmaeili [34] used molecular dynamics finite element method to measure the strength and fracture strain of several straight and helical nanotubes with different diameters under the tensile load. Their findings showed that by increasing the diameter of helical nanotubes, the fracture force is increased, while the fracture strain is constant. Feng et al. [35] evaluated the spring stiffness of a three-turns carbon nanospring around 0.36 N/m and maximum elongation of 38% in elastic deformation. In another research [40] the mechanical responses and distributed partial fractures in single- and multi-strand helical CNTs with toughness up to 5000 J/g by MD simulations of tension tests were reported. Shahini et al. [7] studied the effects of temperature and pitch angle on the tensile properties of CCNTs with different chiral vectors. It was found that by decreasing the rising angle, the yield strength and elastic slope decreases while the yield strain, failure strain, and toughness increase. In a recent study, Wu et al. [10] assessed the role of CNT-chirality in their mechanical performances. They reported that for armchair and zigzag CCNTs, the unusual extensibility is accomplished by well-distributed nanohinge-like plastic deformation, whereas for chiral ones this is contributed by superelasticity and nanohinge-like fracture mechanisms.

In general, it can be concluded that the tensile properties of CCNTs are strongly dependent on the geometry and the chirality of CNTs [7–11]. It should, however, be noted that many questions have remained yet without any answer regarding the mechanical properties of CCNTs. Due to the complex conditions in the tensile test of CCNTs and the infinite number of possible structures, the accurate mathematical expression of mechanical properties as a function of geometrical parameters is not well identified and formulated as yet. Hence, finding structures with excellent mechanical characteristics such as high yield strength and yield strain is not achievable through a process of computational trial-and-error or experimental methods. Moreover, developing accurate theoretical equations between the mechanical properties with each other or with morphological variables such as coil and tube diameter, pitch angle, pitch length, and the symmetry of their top view motifs is a vital factor in the mechanical design of CCNTs. The objective of this work is to employ an efficient multi-objective process optimization framework to find the preeminent structures with respect to their mechanical properties. Furthermore, guided by insights from the multi-objective optimization, a continuum model is fitted to the results of MD simulation for developing several analytical equations. Finally, the detailed explanation of the superelastic mechanisms of small and large CCNTs is discussed.

## 2. Models and Methods

### 2.1. Structural modeling of CCNTs

Systematic modeling of CCNTs as a function of carbon atoms is an intricate graph-theoretical problem because of their nonlinear helical morphology and existence of non-hexagonal carbon rings. Here, we used the generalized construction scheme of helical CNTs proposed by Chuang et al. with some modifications for our purposes [41–45]. Four major steps were carried out to generate a CCNT:

1. It is widely known that imposing non-hexagonal carbon rings into a graphene sheet introduces Gaussian curvature [46,47]. For the first step, the desired polygon TCNT (Figure 1b) was obtained from a planar graphene sheet (Figure 1a) through a cut-and-fold procedure. The widths and heights of the BCC′B′ and ADD′A′ rectangles supply four degrees of freedom to define any possible TCNT. The four indices from Figure 1 ($n_{75}$, $n_{77}$, $n_{55}$, s) are defined as follows:

$n_{75}$ = topological distance between inner-ring heptagons and outer-ring pentagons

$n_{77}$ = topological distance between inner-ring heptagons

$n_{55}$ = topological distance between adjacent heptagons and pentagons along the vertical direction

s = length of the unit cell

2. In the next step, with either inner-rim or outer-rim horizontal shifting (HS), or even a combination of both, distorted TCNTs were created. A 30-degree rotation (HS=1) for the central hollowed hexagonal hole is defined by moving heptagons or pentagons to suitable coordinates on the graphene honeycomb lattice as shown in Figure 1c. With carefully chosen parities of the four indices and the size of horizontal shifting described above, one can obtain TCNTs with $D_{nd}$ or $D_{nh}$ symmetries as shown in Figure 1d,e [45].

3. It should be noted that the achieved TCNTs by step 2 possess high strain energy. By dissecting the TCNT along any of its longitudes, the strain energy is released and a CCNT is generated. The more distorted the initial parent TCNT is, the more helicity the CCNT obtained [45]. The relation between pitch angle and the HS is expressed as [44]:

$$\theta = \arctan\left(\frac{HS}{g}\right); \text{ for } HS < 10 \qquad (1)$$

Where θ is the pitch angle, and g is the circumference or girth of the TCNT. Moreover, the number of atoms in the CCNT is given by

$$N = 2n_{77}.s + 2n_{75}(2s + n_{75}) + 2n_{55}(n_{75} + s) \qquad (2)$$

4. The geometry optimization was carried out by molecular mechanics method with pairwise potential proposed by Lenosky et al. [48] to make the bond length and the bond angles close to 1.43 Å and 120 degrees, respectively.

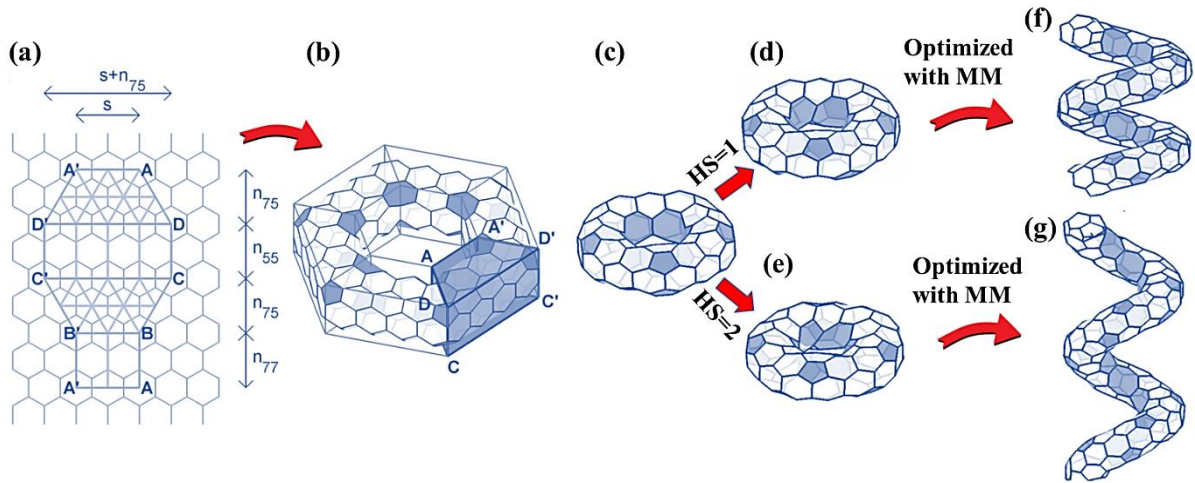

Figure 1. The formation of a CCNT from its parent TCNT. (a) a planar graphene sheet for the cut and fold process. (b) A typical $D_{6d}$ TCNT with indices (n75, n77, n55, s) = (2, 1, 1, 2). The shaded region shown is a particular rotational unit cell. (c) A typical parent $D_{5h}$ TCNT. (d) TCNT after an HS operation. The HS of this TCNT is unity. (e) with HS = 2. (f) CCNT derived from HS-TCNT in part (d). CCNT obtained by dissecting the HS-TCNT shown in previous parts at any of its longitudes. (g) CCNT derived from the HS-TCNT in part (e).

Therefore, six indices suffice to determine a typical CCNT; the first four indices correspond to the parent TCNT, the fifth index is either 1 for $D_{nd}$ symmetry or 2 for $D_{nh}$ symmetry, and the last index specifies the HS parameter. By changing these six indices, all possible CCNTs with different coil diameter, tube diameter, pitch angle, and pitch length can be modeled. it is noteworthy that not every combination of these six indices yields a stable structure. As a consequence, the structures were examined to verify if they were thermodynamically stable before the tension. In the following, we do not strictly distinguish CCNT, Helical CNT, nanocoil, and nanospring, hence they will be used interchangeably.

**2.2. Multi-objective process optimization, Pareto front, and NSGA-II**

A multi-objective optimization problem is an optimization problem that involves multiple objective functions [49]. For the present work, it can be mathematically formulated as

$$Max\ f(a_i) = (f_1(a_i), f_2(a_i))'$$

(3)

$$s.t.\ a_i \in A, i \in \{1,2,\ldots,6\}$$

Where $f_1$ and $f_2$ are the objective or fitness functions which can be yield strength or yield strain, $a_1, \ldots, a_6$ are the six indices needed to identify a CCNT and the set $A$ is the feasible set of decision vectors that will be defined regarding the nanotube size (see section 3). In bi-objective optimization, there does not typically exist a feasible solution that minimizes both objective functions simultaneously [50]. Consequently, attention is paid to Pareto optimal solutions. That is to say, solutions that cannot be improved in any of the objectives without degrading another objective [50]. In mathematical terms, a feasible solution $a^1 \in A$ is said to (Pareto) dominate another solution $a^2 \in A$, if [51]

$$f_i(a^1) \leq f_i(a^2)\ \forall\ i \in \{1,2\} \text{ and,}$$

$$f_j(a^1) \leq f_j(a^2)\ \exists\ j \in \{1,2\} \tag{4}$$

A solution $a^* \in A$ (and the corresponding outcome $f_i(a^*)$) is called Pareto optimal if there does not exist another solution that dominates it. The set of Pareto optimal outcomes is often called the Pareto front or solutions with the first rank [50]. If the solutions related to the current Pareto front are eliminated from the set $A$, the new Pareto front is considered to be the solutions with the second rank. The same procedure can be used to determine other ranks.

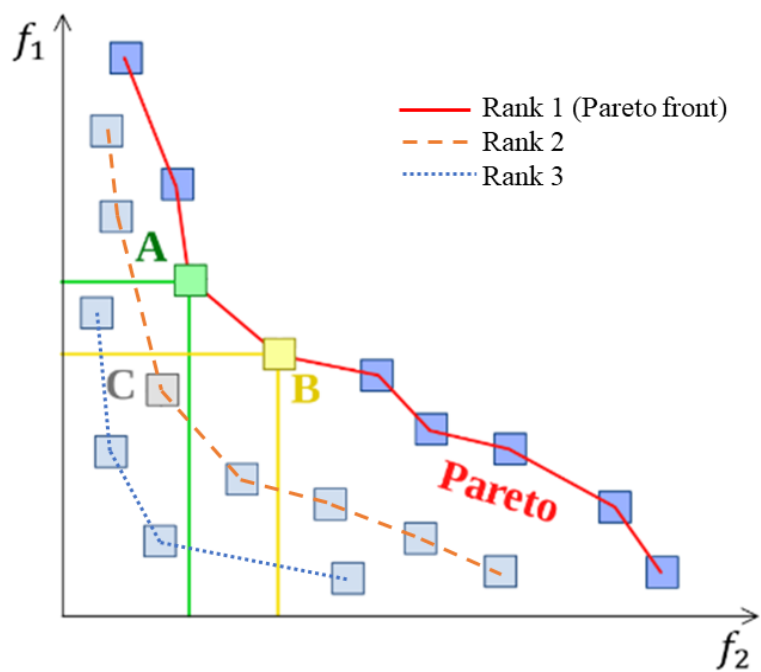

Figure 2. Example of a Pareto front (in red). The boxed points represent feasible solutions, and larger values are preferred to smaller ones. Point C is not on the Pareto frontier because it is dominated by both point A and point B. Points A and B are not strictly dominated by any other, and hence do lie on the Pareto front. The solutions with ranks 2 and 3 are drawn with orange and blue dots, respectively.

There are numerous multi-objective optimization techniques. For this paper, the non-dominated sorting genetic algorithm II (NSGA-II) was used as the optimization algorithm [52]. Crowding distance is used as a second-order sorting criterion. NSGA-II creates and fills a mating pool, using binary tournament selection. Then, crossover and mutation operators are applied to certain portions of the mating pool members. Starting from a random geometrical point, the NSGA-II was iteratively applied. The optimization process was halted when no new point was added to Pareto optimal solutions for 10 iterations.

**2.3. Molecular dynamics simulation**

All calculations were carried out in the LAMMPS molecular dynamics simulation package using the AIREBO potential field [53,54]. The many-body short-range REBO forcefield is capable of modeling the breaking and formation of covalent bonds between carbon atoms during the tensile test. In order to prevent the spurious strain hardening behaviors during tension, the cutoff distance in the switching function of the short-range REBO potential was selected to be 2.0 Å [55]. For the Lennard-Jones potential field, a cut off radius of 10.2 Å was selected to ensure the application of the potential at large distance. A periodic boundary condition (PBC) was adopted to preclude the edge effects along the axial direction of helical CNT and non-PBCs were adopted along two other directions. Before the tensile test, CCNTs were given 50 ps at 300 K to relax in zero bar pressure condition in the NPT (Isothermal-Isobaric) ensemble. The pressure and temperature control of the system were performed by the Nosé-Hoover's barostat and thermostat, respectively [56,57]. Time steps of 0.5 fs and velocity-verlet integration algorithm was adopted to integrate the equation of motions in all simulations. In the tensile simulations, a constant engineering strain rate of $10^9 \ s^{-1}$ was applied. During the tension, the NVT (Canonical) ensemble and Nosé-Hoover thermostat were used. The tensile stress was calculated using the virial equation [58–60]. As suggested by previous studies, the dissociation of the first atomic bond was considered as the elastic limit of helical CNTs [7–9]. Therefore, the elongation was stopped whenever a bond was dissociated.

## 3. Results and discussion

In this section, the yield stress ($\sigma_y$) and yield strain ($\varepsilon_y$) are considered as the objective functions. It is found in our optimizations that CCNTs with large indices dominate the smaller ones due to their superior mechanical properties. To this end, the first four indices which control the size of the CCNTs are divided into two categories. In the first category, the $a_1$ to $a_4$ indices are selected from 1 to 5. This class of CCNTs possesses small tubes and coil diameters. The second category consists of CCNTs with the first four indices in the range of 5 to 9, consequently, the CCNTs are larger especially concerning their tube diameters. It should be pointed out that there is no limitation for the values of the indices, but larger CCNTs increase the computational cost immensely. Additionally, the results can be predicted for larger CCNTs which will be discussed in section 3.3.

### 3.1. Small CCNTs

The results of the structural multi-objective optimization for small CCNTs are shown in Figure 3. Each point represents a distinct CCNT with its corresponding yield point values. The Pareto optimal solutions are illustrated in red circles. This figure is revealing in several ways. First, this curve proposes the optimal nanocoils regarding their yield strength and yield strain with the smallest possible dimensions. With careful choosing the indices, there is a nanohelix that can be elongated up to ε=2.95 in the elastic region. Second, the relation between σ_y and ε_y in the Pareto front can be defined as a power function by the equation $\sigma_y = k\varepsilon_y^n$, where $k$ and $n$ are constants that are around 10 and -1, respectively. In other words, the result indicates that there is a limitation for achieving mechanical properties of helical CNTs, i.e. if one property

($\varepsilon_y$) increases the other ($\sigma_y$) decreases and vice versa. The other feasible solutions for the $\sigma_y$ - $\varepsilon_y$ optimization are shown in blue dots. Most of the solutions are distributed in strains less than 0.75 and stress less than 55 GPa, suggesting finding structures that can resist high strains is unlikely. Fortunately, the NSGA-II algorithm enables us to find those even scarce structures through the crossover and mutation process. The inner plot of Figure 3 presents five stress strain-strain curves of Pareto solutions for different CCNTs. It is readily observed that the stress-strain correlation is almost linear for all kinds of small nanohelixes.

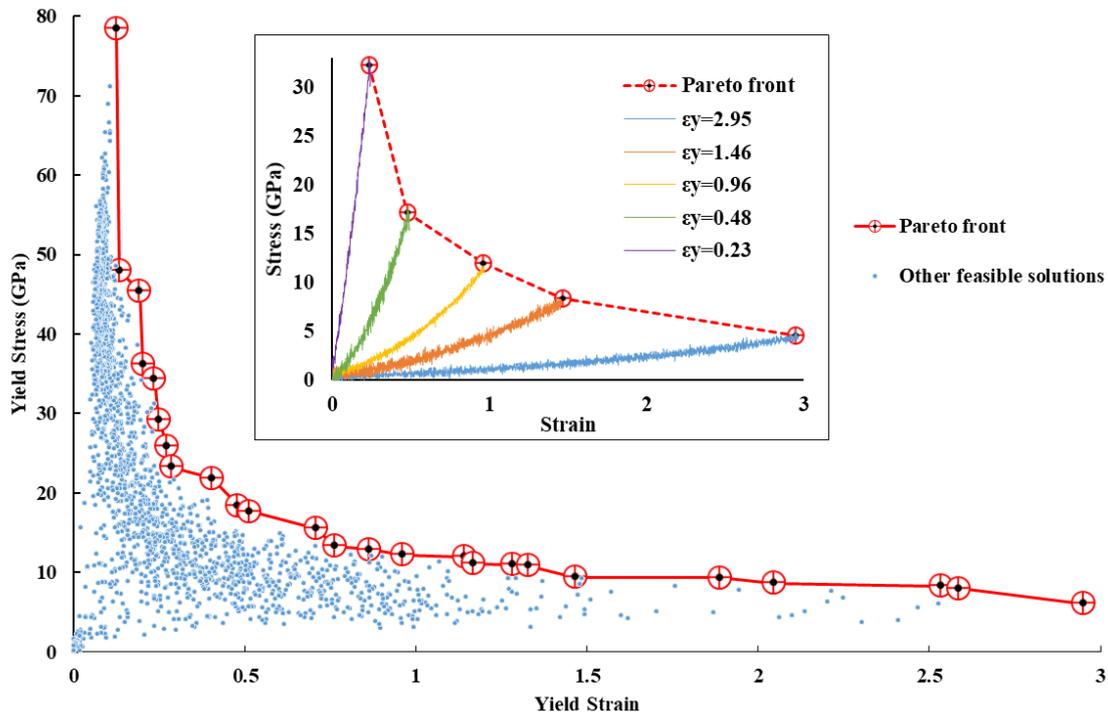

Figure 3. The Pareto optimal (red circles) and other feasible solutions (blue dots) for multi-objective optimization of yield stress vs yield strain for small CCNTs. Each point shows a separate nanocoil with unique indices. Most of the solutions have high yield strength while a few attain superelongation. The inner plot displays the stress-strain curve for five CCNTs from Pareto front with different yield strain.

The snapshots of the CCNTs in the inner plot of Figure 3 are shown in Figure 4 at their yield point. The top-view contours of the atomic stress reveal that for all of the nanotubes the majority of stretching load is absorbed by the inner edges of the CCNTs (Video S1). Since the heptagonal carbon rings are located in this region, in addition to weak binding energy between the carbon atoms in the heptagonal rings, it is more likely that the first bond dissociation occurs in the inner-edge and in heptagonal rings. Conversely, the outer edge of CCNTs is either in compression or low strain concentration.

Careful observation in Figure 4c-e shows that CCNTs with yield strains larger than 1.0 are characterized by a series of buckling mechanisms. This buckling deformation has also been observed experimentally [29] and predicted via MD simulations before [8,40]. For the structures with superelastic behavior, there are also other mechanisms responsible for this unusual behavior such as the formation of kinks (red arrow in Figure 4c) and elastic "nanohinges" (black arrow in Figure 4c), which remarkably remedy the stress concentration. This behavior was predicted only in the plastic region before [40].

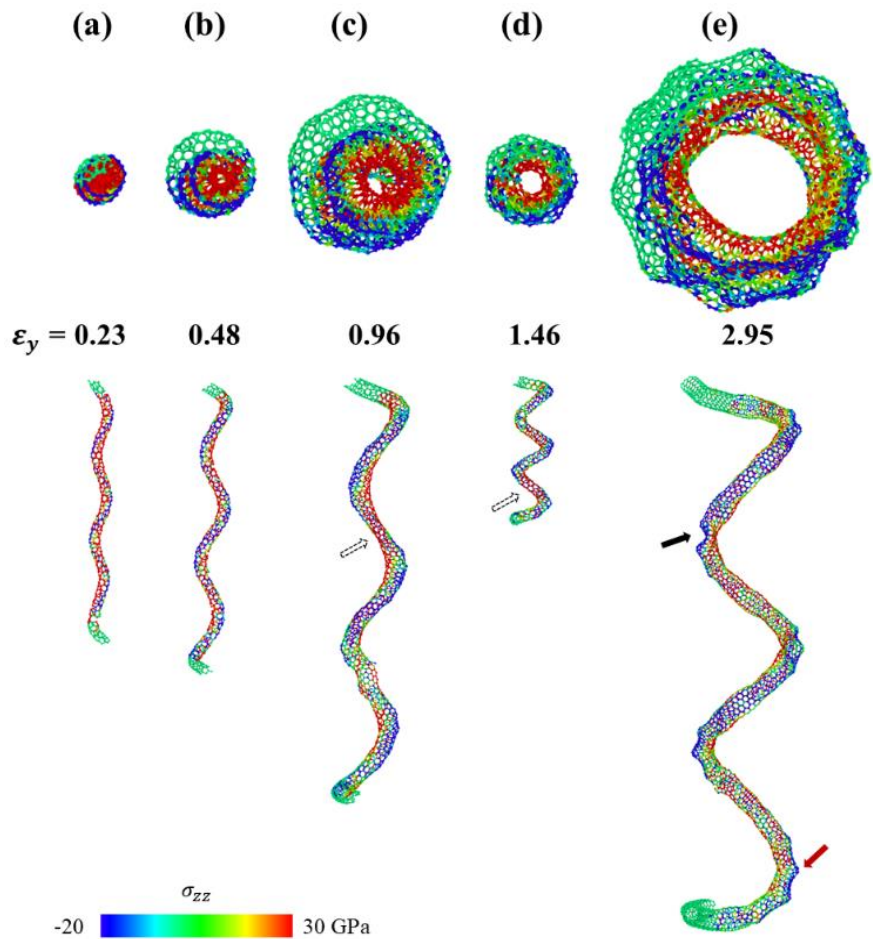

Figure 4. The top- and side-view of the molecular structural configuration of five CCNTs with different yield strain at the yield point. Significant stress concentrations on the inner edge of CCNTs are clearly observed. The arrows indicate the buckling, and formation of kinks and nanohinges. The atoms are colored according to von Mises stress.

Upon closer inspection on top-view snapshots of Figure 4, it is found that the inner diameters of CCNTs with high yield strains are maintained while it approaches zero for low yield strain regimes. Further examination of the inner diameter of CCNTs is depicted in Figure 5 and Video S2. Two CCNTs from Pareto front with similar initial inner coil but different yield strains are selected. The displacement of atoms in grey lines indicates that for helical CNTs with low yield strain, the middle atoms considerably displaced horizontally in the XY plane, while for the other structure, the middle atoms moved in short distances along the load direction (z) thus the nanocoil maintains its coil diameter.

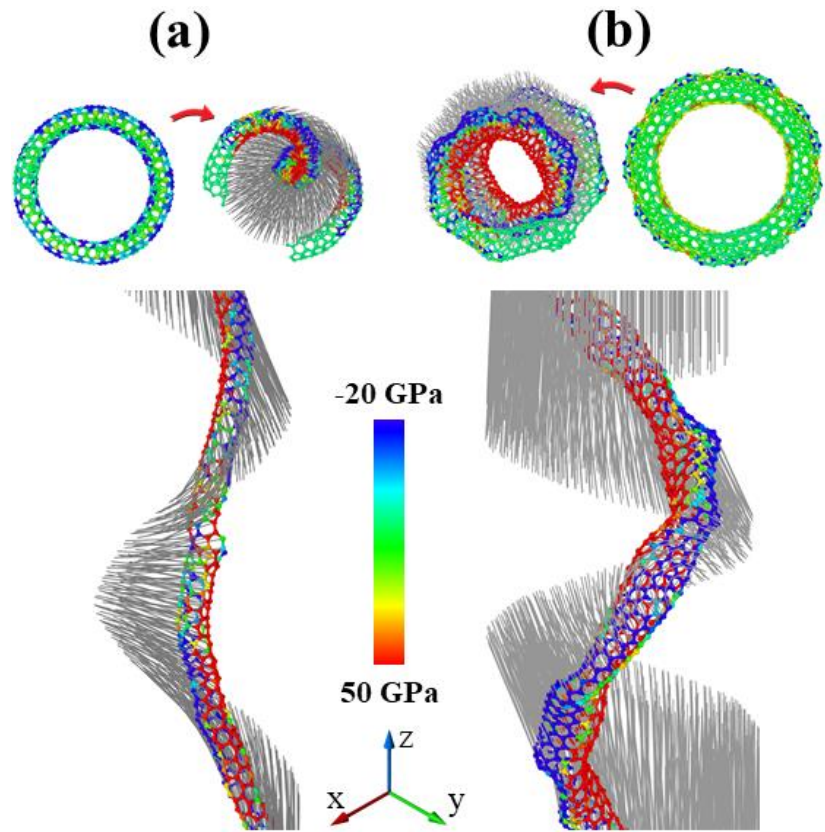

Figure 5. The displacement analysis of middle atoms of two CCNTs with similar initial coil diameter but different yield strain. The grey lines show the displacement vectors of atoms during tension. (a) The nanohelix with low yield strain atoms move horizontally while in (b) the middle atoms of CCNT with high yield strain move in short distances along the load direction. The atoms are colored according to von Mises stress.

### 3.2. Theoretical Model

In order to provide physical insight into the contribution of geometrical parameters on the tensile properties of CCNTs, and to justify the correlation of $\sigma_y$ - $\varepsilon_y$ in Pareto front, it is beneficial to model a CCNT with an equivalent continuum model. As a first-order estimation, a CCNT can be considered as a thin helical bar with the following governing equations [61,62]:

$$\sigma = \frac{32PR\sin\theta}{\pi d^3}(1 + \frac{d}{8R}) \tag{5}$$

$$\tau = \frac{16PR\cos\theta}{\pi d^3}\left(1 + \frac{d}{3R}\right) \tag{6}$$

Where $\sigma$ and $\tau$ are the normal and shear stress, $P$ is the axial load, $R$ is the mean coil radius, $\theta$ is the pitch angle, $d$ is the diameter of the coil wire. For small indices CCNTs, the second term of Eq. (5) which is $\frac{d}{8R}$ can be neglected. Therefore, the maximum principal stress can be obtained by:

$$\sigma_1 = \frac{\sigma}{2} + \sqrt{\left(\frac{\sigma}{2}\right)^2 + \tau^2} \tag{7}$$

As a result,

$$\sigma_1 = \frac{16PR}{\pi d^3}(1 + \sin\theta) \tag{8}$$

Using thermoelastic analysis, it has been numerically shown that appropriately averaged (spatial and temporal) virial stress is the Cauchy stress [63]. Figure 6 shows the six stress components in the tensile simulation of two different CCNTs with low and high pitch angles. Surprisingly, unlike the macroscale engineering springs where the shear stress has the most contribution to the stress tensor [62], normal stress in the load direction ($\sigma_{zz}$) is the only stress component that controls the tensile behavior of nanosprings. As a consequence,

$$\sigma_1 = \sigma_{zz} \tag{9}$$

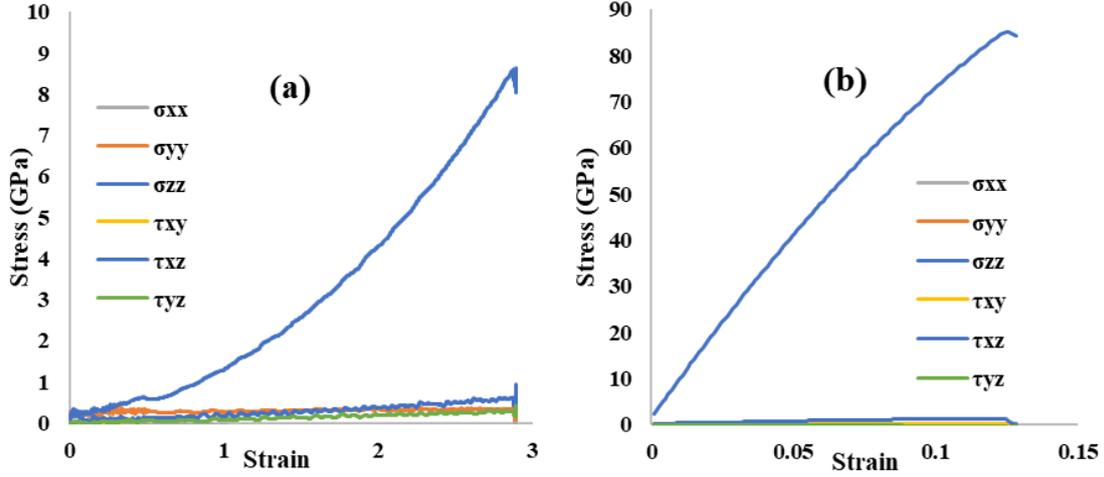

Figure 6. Variations of six stress components in stress-strain curves for (a) CCNT with the pitch angle = 10 degrees and (b) pitch angle = 85 degrees. The domination of $\sigma_{zz}$ in stress tensor for the uniaxial test of helical CNTs is obvious.

The total strain in the axial direction for an open-coil spring is calculated by [62]:

$$\varepsilon = \frac{64PR^3 l}{d^4 \cos\theta}\left(\frac{\cos^2\theta}{G} + \frac{2\sin^2\theta}{E}\right) \qquad (10)$$

Where $l$, $G$, and $E$ are the initial pitch length of the CCNT, shear and elastic modulus of a CNT, respectively. Substituting the $P$ from Eq. (10) into the Eq. (8), one has:

$$\sigma_{zz} = \frac{d.l.\xi}{4\pi R^2} \cdot \varepsilon_{zz} \qquad (11)$$

Where $\xi$ is a function of pitch angle,

$$\xi = \frac{E.G.\cos\theta(1+\sin\theta)}{E.\cos^2\theta + 2G.\sin^2\theta} \qquad (12)$$

From Eq. (11), it can be concluded that CCNTs with high yield strain are characterized with low pitch length, tube diameter, pitch angle, and high coil radius. To shed light on the relation of Eq. (11) and MD results, all the optimal structures of small CCNTs are displayed in Figure 7. It can be seen that structure with the lowest yield strain resembles straight CNT whereas structures with high yield strain are close-coil nanosprings. Generally, as the yield strain increases, the coil diameter initially increases, then decreases, and finally increases again. As we look at the top-view of CCNTs from left to right in Figure 7, whenever the coil diameter reduces, the other geometrical parameters (e.g. $d, l, \theta$) reduce as well to compensate for the reduction of coil diameter thus raising the yield strain. That is, the amount of these four geometrical variables determine the yield points values.

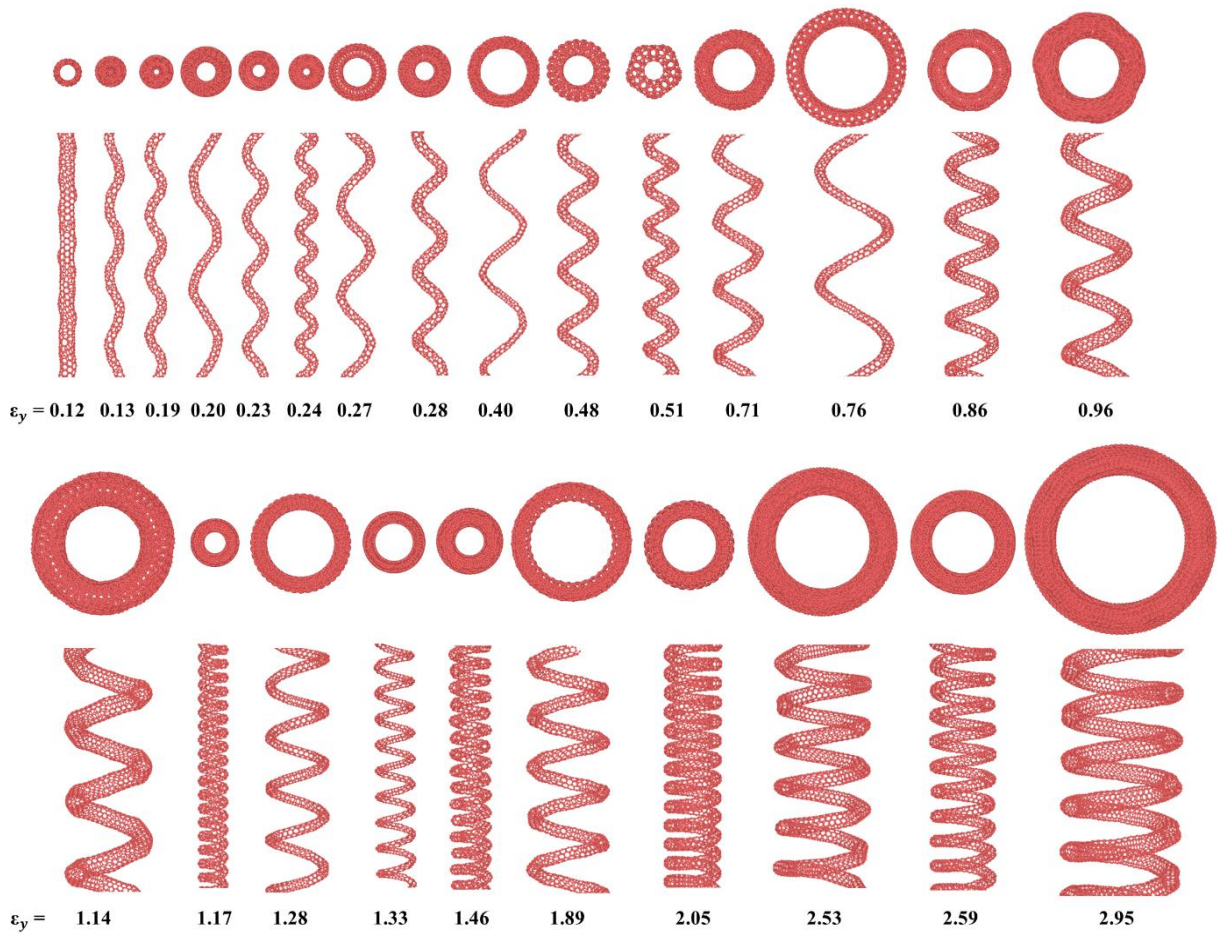

Figure 7. The atomic structural of Pareto optimal solutions for small CCNTs. The yield strain increases from left to right. Overall, by increasing the yield strain the coil diameter increases while the pitch angle decreases.

If we consider the tensile behavior of small nanocoils linear in the elastic region, it can be inferred from Eq. (11) that the elastic modulus of a CCNT is a function of its geometrical parameters. These parameters are detailed in

Table 1 for Pareto front solutions of small CCNTs. The elastic modulus was calculated from the continuum equation ($E_{continuum}$) and MD simulations ($E_{MD}$) and compared in

Table 1. Interestingly, for pitch angles less than 35 degrees, there is a satisfactory agreement between the simulation results and the Eq. (11). However, as the pitch angle increases, the difference between E$_{MD}$ and E$_{continuum}$ becomes larger. This is because at high pitch angles $d$ approaches $R$, therefore, the $\frac{d}{8R}$ term in Eq. (5) is no longer negligible. To deal with this problem, a new coefficient which is a function of pitch angle is introduced to the Eq. (11). As a result, the modified stress-strain equation in the elastic region of small CCNTs can be obtained by

$$\sigma_{zz} = \frac{d.l.\xi.k}{4\pi R^2} \cdot \varepsilon_{zz} \tag{13}$$

Where $k$ is the upper mentioned coefficient and obtained by fitting the continuum model to the MD simulation results,

$$k = 0.73\, e^{1.254\theta} \tag{14}$$

Table 1. Structural parameters and modulus of elasticity for Pareto optimal solutions of small CCNTs.

| Index | d (Å) | l (Å) | R (Å) | θ (degree) | E$_{continuum}$ (GPa) | E$_{MD}$ (GPa) | E$_{modified}$ (GPa) |
|---|---|---|---|---|---|---|---|
| (2,5,2,5,2,1) | 6.48 | 37.00 | 22.26 | 10 | 1.94 | 1.84 | 1.76 |
| (1,3,2,2,2,1) | 4.41 | 16.17 | 10.01 | 10 | 2.85 | 2.89 | 2.59 |
| (2,1,2,4,1,1) | 4.72 | 21.36 | 11.64 | 10 | 2.98 | 3.09 | 2.71 |
| (2,3,2,5,2,1) | 6.25 | 38.55 | 16.96 | 11 | 3.40 | 3.26 | 3.16 |
| (1,2,1,1,1,1) | 3.28 | 16.95 | 4.87 | 13 | 9.71 | 9.49 | 9.43 |
| (2,2,1,5,1,1) | 4.46 | 52.49 | 14.47 | 15 | 4.76 | 4.94 | 4.82 |
| (1,1,1,2,1,1) | 2.56 | 25.37 | 6.96 | 15 | 5.70 | 6.85 | 5.78 |
| (2,1,1,2,1,1) | 4.08 | 14.91 | 6.64 | 16 | 5.93 | 6.40 | 6.15 |
| (1,2,2,3,1,1) | 4.00 | 58.87 | 11.65 | 19 | 7.67 | 8.54 | 8.50 |
| (3,2,4,5,1,2) | 8.45 | 67.84 | 15.24 | 21 | 11.12 | 10.48 | 12.86 |
| (1,3,2,2,1,1) | 4.70 | 47.20 | 9.01 | 22 | 12.39 | 12.16 | 14.65 |
| (3,1,2,5,1,1) | 6.04 | 62.46 | 12.02 | 23 | 11.95 | 12.68 | 14.44 |
| (2,1,1,4,1,1) | 3.86 | 59.91 | 8.61 | 28 | 14.73 | 21.93 | 19.85 |
| (1,1,3,4,1,1) | 4.17 | 125.89 | 14.16 | 31 | 12.52 | 17.49 | 18.02 |
| (2,1,1,2,1,2) | 4.16 | 35.73 | 5.27 | 32 | 25.65 | 34.46 | 37.74 |
| (1,2,2,2,2,1) | 4.02 | 52.17 | 6.51 | 35 | 23.87 | 38.40 | 37.50 |
| (1,1,1,4,1,1) | 2.67 | 110.94 | 8.23 | 40 | 21.06 | 41.71 | 36.92 |
| (1,2,2,2,1,1) | 4.10 | 63.48 | 5.01 | 45 | 49.10 | 81.71 | 96.02 |
| (1,2,1,1,2,2) | 3.60 | 32.39 | 2.96 | 47 | 62.26 | 118.00 | 127.20 |
| (1,2,1,3,1,1) | 3.29 | 93.73 | 6.06 | 48 | 39.03 | 83.00 | 81.51 |
| (1,2,1,2,2,1) | 3.46 | 53.75 | 3.70 | 53 | 59.88 | 145.83 | 139.51 |
| (1,1,2,4,1,1) | 3.22 | 116.36 | 4.88 | 58 | 64.37 | 178.70 | 167.30 |
| (1,2,1,2,1,2) | 2.90 | 59.46 | 2.75 | 59 | 91.70 | 236.81 | 243.62 |
| (1,2,1,3,1,2) | 3.64 | 52.01 | 2.41 | 64 | 116.72 | 361.74 | 345.95 |
| (1,1,1,5,1,7) | 5.80 | 100.19 | 1.92 | 85 | 124.79 | 613.44 | 585.65 |

Figure 8 and Table 1 suggest that analytical equations appear to be well substantiated by the correction factor. However, careful attention must be paid in using Eq. (13) since it only applies to CCNTs that are in Pareto front or in the solutions with the rank of less than 8. Furthermore, for having superelasticity in a nanohelix, owning high coil diameter and low pitch angle is necessary but not sufficient. The arrangement of non-hexagonal defects, especially the position of heptagonal carbon rings which absorb the most amount of tensile force, is another factor to be considered. From Table 1 we can find 19 CCNTs with $D_{nd}$ symmetry against only 6 structures with $D_{nh}$ symmetry. Hence structures with $D_{nd}$ symmetry in their parent TCNT are preferred for small nanocoils.

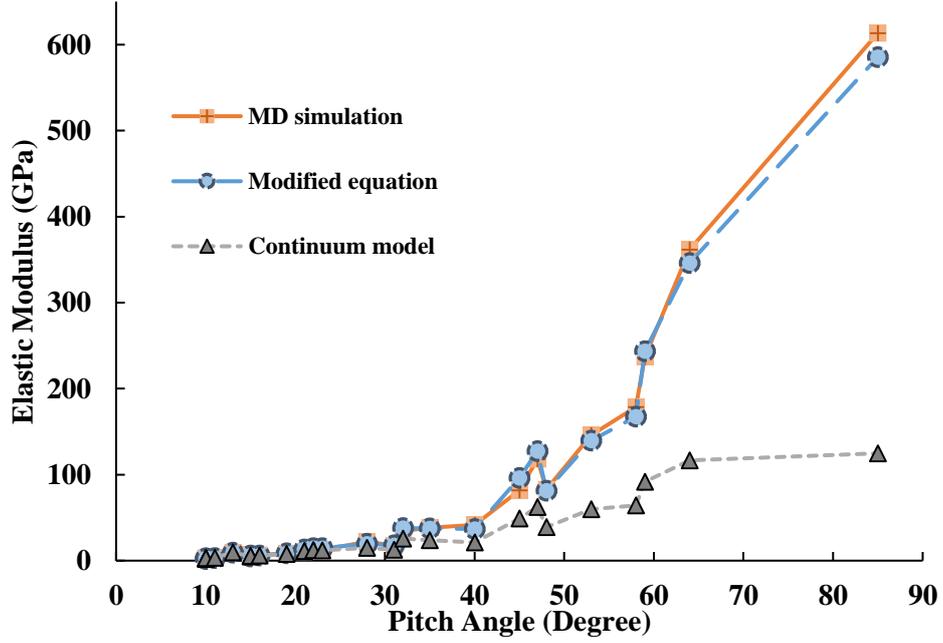

Figure 8. Comparison of the elastic modulus calculated by MD simulations, Eq. (11), and modified Eq. (13) proposed in this work. By using the appropriate coefficient, the tensile properties of CCNTs can be expressed as an analytical equation.

As stated previously the correlation between $\sigma_y$ and $\varepsilon_y$ in the Pareto front solutions can be formulated mathematically by fitting a power function as shown below

$$\sigma_y = 10\varepsilon_y^{-1} \qquad (15)$$

Solving Eq. (13) at the yield point and substituting it into Eq. (15), one has

$$\varepsilon_y = \frac{2R\sqrt{10\pi dl\xi k}}{dl\xi k} \qquad (16)$$

$$\sigma_y = \frac{\sqrt{10 dl\xi k}}{2\sqrt{\pi} R} \qquad (17)$$

Eqs. (16) and (17) express the yield stress and yield strain as a function of geometrical parameters and can be used to calculate the highest possible $\sigma_y$ and $\varepsilon_y$ one can obtain in small CCNTs. We observe from Figure 9 that apart from a slight discordance for strains higher than 2.0, the predicted results from analytical equations are in appreciable agreement with MD results. The prime cause of the discrepancy is the chosen fitting function for the Pareto front. for the sake of simplicity in developing the equations, we used -1 instead of -0.86 for the power of ε in Eq. (15).

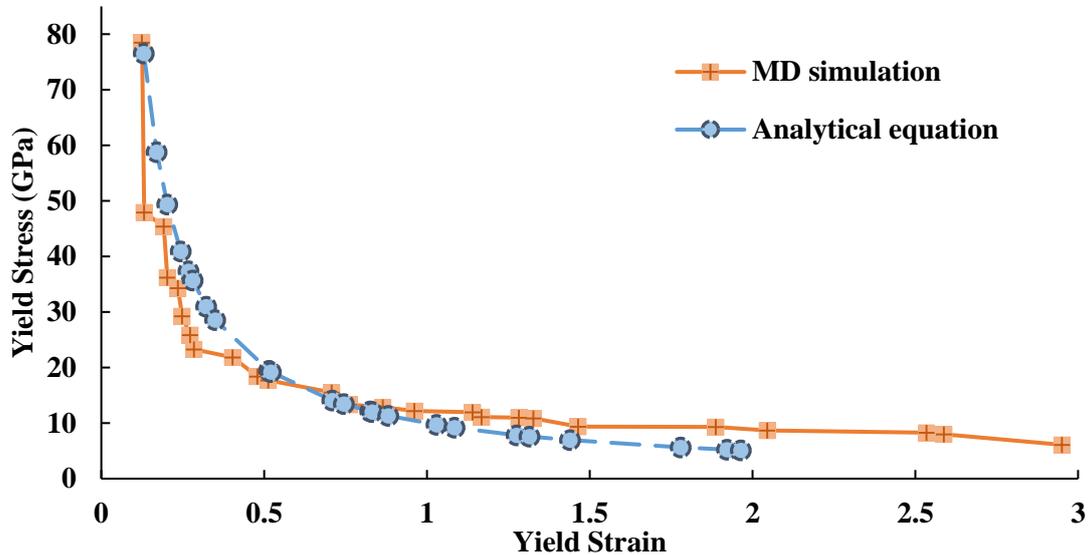

Figure 9. The Pareto front resulting from MD simulations in blue points versus the predicted Pareto front from Eqs. (16) and (17). The apparent lack of correlation in large strains can be attributed to the simplicity of fitting function.

### 3.3. large CCNTs

The results of the multi-objective optimization of large nanocoils are shown in Figure 10. In this figure, structures with superelastic behavior are found that can be stretched up to four times of their initial length. As far as we know, no one has predicted these amounts of elongation in the elastic region. Similar to small CCNTs, the Pareto front can be fitted to a power function but with higher $k$ and lower $n$ than small ones. To note the similarities and differences between the small and large CCNTs, the Pareto front of small nanocoils is added to Figure 10. It can be seen that the Pareto optimal solutions of large CCNTs have relatively high amounts of yield point values as compared to small CCNTs. This suggests that with increasing the indices and the size of the nanotube, the mechanical performance improve. Another optimization for nanotubes with indices from 1 to 9 was investigated and the results were identical to Pareto optimal solutions of large CCNTs. This means there is no combination of small and large indices that can dominate the CCNTs with large indices.

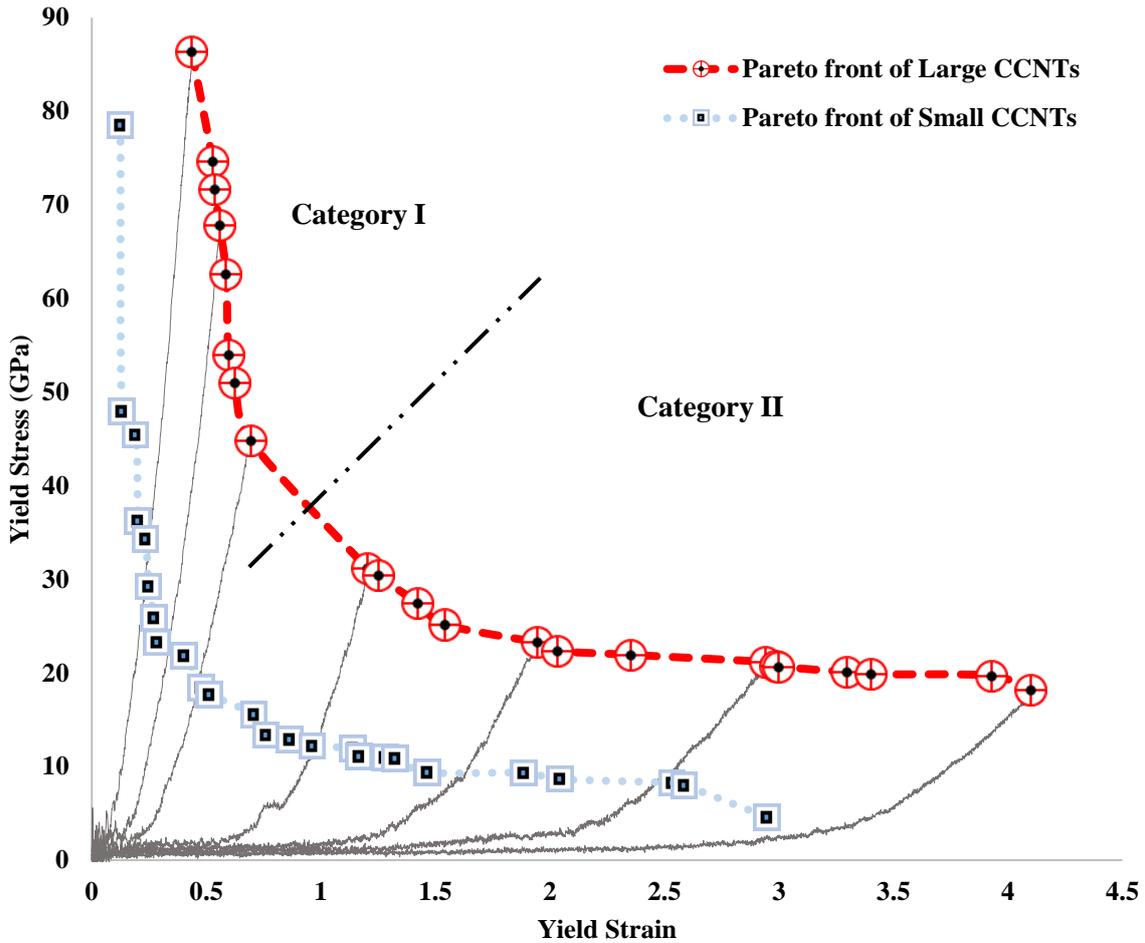

Figure 10. The Pareto front for multi-objective optimization of large CCNTs (in red circles). For the sake of comparison, the Pareto front of small CCNTs also presented in the blue squares. Since the yield point values of large CCNTs are higher than small ones, it can be concluded that the mechanical properties improve as the indices increases.

Figure 11 shows the molecular configurations of Pareto optimal solutions for large CCNTs. It displays a clear trend in the structural parameters as the yield strain increases; As expected and validated by the continuum model, by increasing the yield strain the radii of coil increases while the pitch length and pitch angle decrease. The tube diameter is almost constant in all structures. The geometrical parameters, yield strength, and yield strain of all the Pareto front solutions for large CCNTs are listed in Table 2. From the first column, one can conclude that the first index is between 5 and 7 while the second index is always 5 indicating the optimal distance for the heptagonal carbon rings should be 5 which is approximately equal to 8.1 Å. There is no general trend in the third index but the fourth index which stands for the segment length is either 8 or 9 which is in the range of 28.8 Å to 31.1 Å. Unlike the small CCNTs, in the large nanohelixes, the $D_{nh}$ symmetry dominates the $D_{nd}$ symmetry and most of the Pareto solutions are from the former symmetry type. The last index which is responsible for the pitch angle reduces by increasing the yield strain.

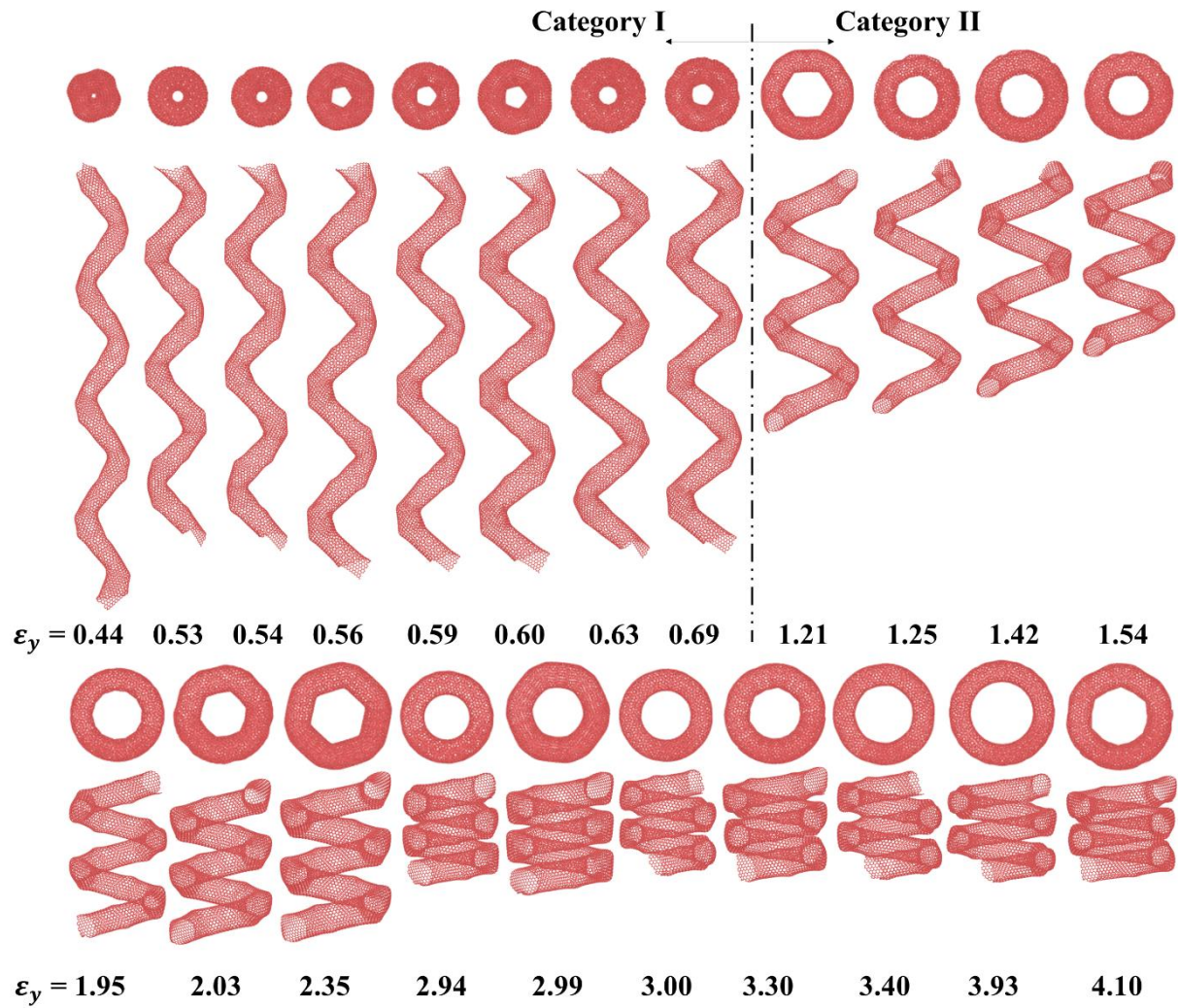

Figure 11. The Molecular configuration of Pareto optimal solutions for large CCNTs. The yield strain increases from left to right. The coil diameter increases and the pitch angle decreases as the yield strain increases. The morphological transformation between CCNTs with $\varepsilon_y$=0.69 and $\varepsilon_y$=1.21 split the structures into two different categories.

Table 2. Structural parameters and corresponding yield point values for Pareto optimal solutions of large CCNTs.

| Index | $d$ (Å) | $L$ (Å) | $R$ (Å) | $\theta$ (degree) | Yield Strain | Yield Stress (GPa) | Category |
|---|---|---|---|---|---|---|---|
| (5,5,5,9,2,9) | 17.54 | 178.66 | 10.58 | 56 | 0.44 | 86.36 | |
| (5,5,6,8,2,9) | 15.95 | 177.03 | 12.84 | 46 | 0.53 | 74.63 | |
| (5,5,7,8,2,9) | 18.00 | 180.07 | 13.01 | 48 | 0.54 | 71.71 | |
| (5,5,8,9,2,8) | 17.55 | 193.57 | 15.65 | 46 | 0.56 | 67.81 | |
| (5,5,7,9,2,7) | 16.64 | 188.24 | 15.61 | 44 | 0.59 | 62.59 | I |
| (6,5,8,9,2,9) | 19.98 | 187.83 | 17.26 | 43 | 0.60 | 53.95 | |
| (7,5,5,9,2,9) | 19.37 | 181.83 | 16.86 | 40 | 0.63 | 50.95 | |
| (6,5,7,9,2,8) | 18.20 | 182.92 | 17.15 | 40 | 0.69 | 44.76 | |
| (5,5,8,9,2,4) | 15.82 | 147.40 | 24.80 | 25 | 1.21 | 31.11 | |
| (5,5,5,9,2,3) | 14.45 | 141.90 | 22.92 | 26 | 1.25 | 30.38 | |
| (6,5,6,9,2,4) | 15.18 | 131.85 | 24.41 | 23 | 1.42 | 27.44 | |
| (6,5,6,8,2,4) | 15.19 | 109.42 | 23.15 | 20 | 1.54 | 25.15 | |
| (5,5,8,8,2,3) | 14.73 | 99.90 | 25.55 | 16 | 1.95 | 23.23 | |
| (7,5,8,8,2,4) | 17.52 | 92.34 | 25.69 | 14 | 2.03 | 22.36 | |
| (7,5,9,9,2,3) | 18.85 | 94.95 | 28.56 | 13 | 2.35 | 21.93 | II |
| (6,5,9,7,1,2) | 16.13 | 62.79 | 24.90 | 12 | 2.94 | 21.16 | |
| (7,5,9,8,1,3) | 17.63 | 68.40 | 27.49 | 11 | 2.99 | 20.74 | |
| (5,5,9,7,1,1) | 15.17 | 70.08 | 25.64 | 12 | 3.00 | 20.65 | |
| (6,5,9,8,2,2) | 17.15 | 71.00 | 27.12 | 10 | 3.30 | 20.09 | |
| (5,5,9,8,1,2) | 15.90 | 75.42 | 28.53 | 11 | 3.40 | 19.85 | |
| (5,5,8,9,1,1) | 16.00 | 68.46 | 30.24 | 9 | 3.93 | 19.70 | |
| (6,5,8,9,2,2) | 16.59 | 67.09 | 29.13 | 9 | 4.10 | 18.14 | |

From Figure 10, Figure 11, and Table 2 it can be observed that there is a gap between CCNTs with $\varepsilon_y$=0.69 and $\varepsilon_y$=1.21 that causes a morphological transfiguration. This change in configuration separates the structures in two different categories: First, the structures with high yield strain that possess high coil radius and low pitch angles. Second, CCNTs with high yield stress which are characterized by low coil radius and high pitch angles. For further investigation of these two types of structures, the stress-strain curves of several CCNTs are exhibited in Figure 10. Referring to this figure, the stress-strain behavior of CCNTs from category I is linear in most part of the tension whereas the pulling stress of other category follow a simple power-law function with $k\varepsilon^n$ scaling, where $k$ is a constant proportional to the elastic modulus and $n$ is a constant depending on the geometry.

Overall, the elastic region of all CCNTs from the Pareto front can be divided into three distinct stages. In the first stage, the elastic slope is small and linear, hence the nanocoil can be elongated at relatively low stretching loads. This low-strain stage ceases whenever the stress increases to a critical amount of 3.5 GPa. This stage has the most contribution to the elongation of CCNTs with high yield strains while it becomes insignificant for nanocoils with low yield strains. A sequence of snapshots of two nanohelixes from both categories is shown in Figure 12 and Video S3. For the first category, this stage is transient but for the second category, this stage contains sequences of vital morphological transformation. First, for the first seven nanocoils with the highest yield strain, because of their small intercoil distance, there exists intercoil van der Waals (vdW) force adhesion that plays a role in the initial elastic loading behavior. The vdW forces cause the reorientation of the coils to follow without any immediate coil separation. With further extensions, the lower turn of the CCNT decoils and the circular cross-section of the tube becomes flattered. The other coil is intact until the strain increases to 0.85 and the coil flattening also occurs for this coil. Consequently, all turns are flattened at the strain of 2.13. From this point, the stretching mechanism is the displacement of atoms towards the center in the longitude direction, thus reducing the coil diameter considerably. This stage is halted after the stress reaches 3.5 GPa. Interestingly, all of these structural transformations occur in relatively low stress where the stress-strain relation is linear. The snapshots of this stage are shown in Figure 12a and d-g.

With further increase in strain, the second stage is commenced. This stage is characterized by the nonlinear increase of tensile stress and a crucial morphological transformation. The displacement of carbon atoms toward the center in the previous stage leads to the generation of a "straight CNT" like fragment in the inner-edge of the CCNT. Hence, the CCNT can be considered as an almost straight CNT with a helical graphene ribbon twisted around it. This stage appears in both categories but lasts longer for nanocoils with high yield strains (Figure 12b and h). In the last stage, the stress increases linearly again but with a higher slope compare to the first stage. This is because the straightening of the inner straight CNT causes significant stress concentrations on the inner-edge of the CCNT and can be observed in Figure 12c and i.

This stage extends until a fully straight CNT generates in the inner part of CCNT and the first atomic bond breaks. By comparison, tensile stiffnesses of nanosprings in this stage are analogous to those of experimentally synthesized ones with large coil radius [27,30,64].

The phase transformations in these three stages account for the superelasticity of these materials while the generation of CNT like fragment and its stability during tension is responsible for high yield stress of large CCNTs. To the best of our knowledge, this is the first time that these kinds of phase transformation are predicted in the elastic region. In fact, CCNTs that are not optimized regarding their structures will yield before they undergo the upper mentioned structural transformations.

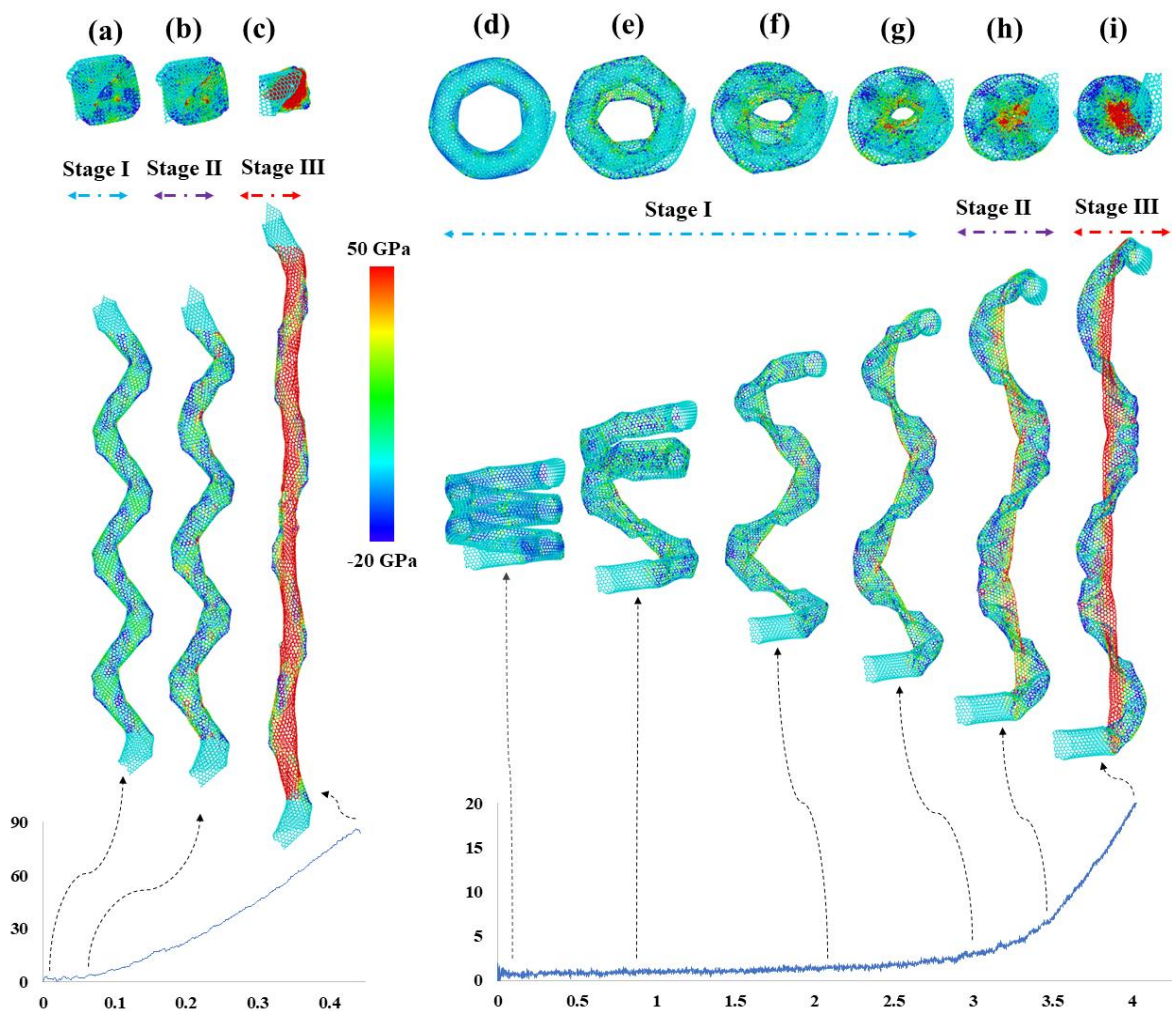

Figure 12. Molecular structural evolution of two large CCNT from the Pareto front. (a)-(c) the CCNT with the lowest yield strain from category I and (b)-(i) the CCNT with the highest yield

strain from category II with their corresponding stress-strain curves. Atoms are colored on the basis of von Mises stress.

Similar to small nanocoils, the analytical equations in the elastic region of large CCNTs is developed by fitting the continuum model to the results of MD simulations. The yield stress for Pareto optimal solutions resulted from MD simulation and Eq. (13) but with a different $k$ are compared in Figure 13a. It can be seen that after the correction factor, our formula reproduces the response of CCNTs in the elastic region. Similar to small CCNTs, the difference between analytical Eq. (11) and MD results increases by the increase in pitch angle. Figure 13b compares the Pareto front resulted from MD simulation and analytical equations similar to Eqs. (16) and (17) after fitting the Pareto front with an appropriate power function. Analogous to small CCNTs, the results are well consistent with MD results except for nanocoils with high yield strains. The results of the multi-objective optimization of small and large CCNTs can be used to predict the larger CCNTs with indices from 10 to 15 and are exhibited with a green dash line in Figure 13. Refer to supplementary information for more details on the correction factor and the corresponding equations for yield stress and yield strain as a function of geometrical parameters. Further studies on the multi-objective optimization of CCNTs regarding their ultimate strain and toughness would be interesting and are currently underway in our research group.

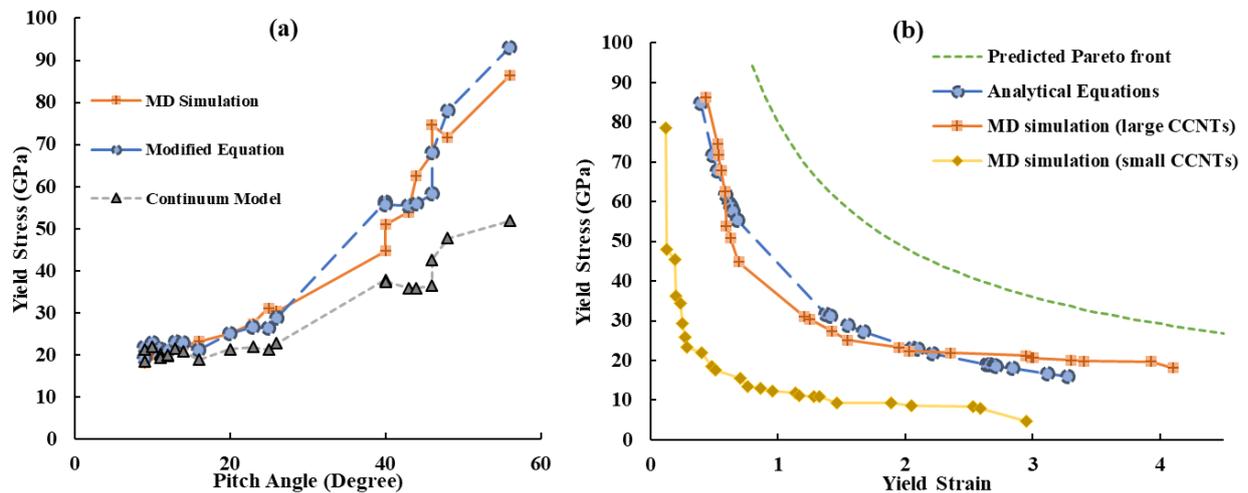

Figure 13. (a) Comparison between simulation and analytical equations of the yield stress as a function of pitch angle for large CCNTs. The modified equation and MD results are in a satisfactory agreement. (b) The Pareto front from both MD and analytical equations for small and large CCNTs and also predicted Pareto front for CCNTs with indices ranging from 10-15.

## 4. Conclusions

Nanoscale helical CNTs have unique mechanical, thermal, and electronic properties that make them suitable for nanoelectromechanical systems. This work focuses on employing a multi-objective process optimization framework for optimizing multiple mechanical properties (e.g. yield strength and yield strain) of small and large CCNTs with respect to their geometrical parameters such as coil and tube diameter, pitch angle, pitch length, and the symmetry of their top view motifs. The multi-objective optimization results show a reverse relation between yield strength and yield strain which can be fitted to a power function by the equation $\sigma_y = k\varepsilon_y^n$ where $n$ and $k$ are constants and depend on the size of CCNT. It is found that by increasing the dimension of CCNTs, mechanical performance improves. The results also confirm that the stretching characteristics of CCNTs are strongly dependent on the geometry. Several theoretical equations are proposed based on fitting a continuum model to the results of reactive MD simulation. The analytical equations can capture the tensile properties of CCNTs in the elastic region. Moreover, A few CCNTs with excellent stretchability in the elastic region are identified. For small CCNTs, the superelasticity of nanocoils in Pareto optimal solutions is attributed to maintaining the inner coil diameter while for large CCNTs the creation of a straight CNT in the inner-edge with a helical graphene ribbon twisted around the CNT-like structure is responsible for remarkable elongations.

## Acknowledgments

The authors would like to thank the research boards at Sharif University of Technology, Tehran, Iran for the provision of the research facilities used in this work. We would also like to show our gratitude to Dr. Javan for sharing his pearls of wisdom with us during the course of this research.

## Appendix A. Supplementary data

Supplementary data related to this article can be found at *DOI…*